\documentclass[12pt]{iopart}

\usepackage{iopams} 
\usepackage{epsfig}
\usepackage{graphicx,psfrag,xspace}

\begin{document}

\title[]{Branching exponential flights: travelled lengths and collision statistics}

\author{Andrea Zoia, Eric Dumonteil, Alain Mazzolo, Sameh Mohamed}
\address{CEA/Saclay, DEN/DM2S/SERMA/LTSD, B\^at.~470, 91191 Gif-sur-Yvette Cedex, France}
\ead{andrea.zoia@cea.fr}

\begin{abstract}
The evolution of several physical and biological systems, ranging from neutron transport in multiplying media to epidemics or population dynamics, can be described in terms of branching exponential flights, a stochastic process which couples a Galton-Watson birth-death mechanism with random spatial displacements. Within this context, one is often called to assess the length $\ell_V$ that the process travels in a given region $V$ of the phase space, or the number of visits $n_V$ to this same region. In this paper, we address this issue by resorting to the Feynman-Kac formalism, which allows characterizing the full distribution of $\ell_V$ and $n_V$ and in particular deriving explicit moment formulas. Some other significant physical observables associated to $\ell_V $ and $n_V$, such as the survival probability, are discussed as well, and results are illustrated by revisiting the classical example of the rod model in nuclear reactor physics.

\end{abstract}

\maketitle

\section{Introduction}

\begin{figure}[t]
\centerline{ \epsfclipon \epsfxsize=9.0cm
\epsfbox{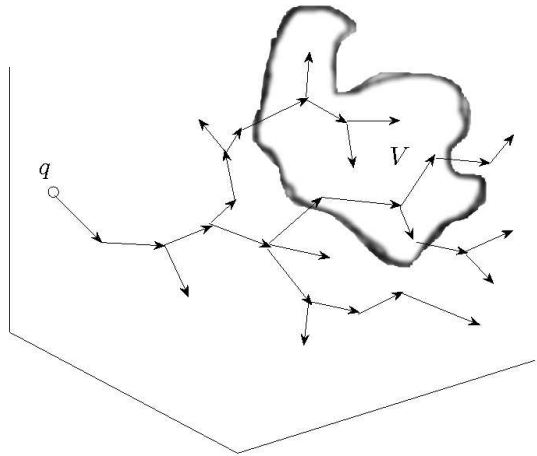} }
\caption{An illustration of branching exponential flights starting from a point source $q$ and traversing a volume $V$ in phase space.}
\label{fig1}
\end{figure}

Consider a single walker initially emitted from a point source at time $\tau_0=0$ at position $\mathbf{r}_0$, with velocity $\mathbf{v}_0$. Once emitted, the walker undergoes a sequence of displacements (at constant speed), separated by collisions with the surrounding medium. When the scattering centers encountered by the travelling particle are spatially uniform, the inter-collision lengths are exponentially distributed~\cite{hughes, weiss}, so that the displacements from ${\mathbf r'}$ to ${\mathbf r}$ in direction $\boldsymbol{\omega}=\mathbf{v}/|{\mathbf v}|$ between any two collisions obey the probability density
\begin{equation}
T({\mathbf r}' \to {\mathbf r}|\boldsymbol{\omega})=\sigma({\mathbf r'},v)e^{-\int_{0}^{\boldsymbol{\omega}\cdot ({\mathbf r}-{\mathbf r'})}\sigma({\mathbf r'}+s\boldsymbol{\omega},v)ds},
\label{exp_kernel}
\end{equation}
with $v=|{\mathbf v}|$~\cite{spanier, lux}. The quantity $\sigma({\mathbf r},v)$ represents the interaction rate per unit length and takes the name of total cross section: $\sigma({\mathbf r},v)$ typically depends on the particle position and speed, and is proportional to the probability of particle-medium interaction along a straight line, carrying units of the inverse of a length~\cite{spanier}. At each collision, the incident particle disappears, and $k$ particles (the descendants) are emitted with probability $p_k({\mathbf r},v)$, whose velocities are randomly redistributed in angle and intensity according to a given probability density $C_k({\mathbf v}' \to {\mathbf v} |{\mathbf r})$, which in principle can vary as a function of the number $k$ of descendants~\cite{spanier}. Each descendant will then behave as the mother particle, and undergo a new sequence of displacements and collisions, giving thus rise to a branched structure, as illustrated in Fig.~\ref{fig1}. As a particular case, when $p_k=\delta_{k,1}$ we recover the well known Pearson random walk~\cite{hughes, weiss}.

Branching random flights as described above lie at the heart of physical and biological modeling~\cite{harris, jagers}, and are key to the description of neutron transport in multiplying media and nucleon cascades~\cite{pazsit}, evolution of biological populations~\cite{lawson}, diffusion of reproducing bacteria~\cite{golding}, and mutation-propagation of genes~\cite{sawyer}, just to name a few. For a detailed survey, ranging from the pioneering work by Galton and Watson on the extinction probability of birth-death processes to the recent developments, see, e.g.,~\cite{harris, pazsit}.

A central question for random walks is to determine the occupation statistics of the stochastic paths in a given portion $V$ of the phase space~\cite{condamin_benichou, condamin, grebenkov, berezhkovskii, agmon_lett, agmon, blanco, mazzolo, benichou_epl, zoia_dumonteil_mazzolo, zdm_prl, zdm_pre}. For exponential flights, the two natural observables of the system are the number $n_V$ of occurred visits to the volume $V$ and the total length $\ell_V$ travelled in $V$~\cite{spanier, lux, blanco, mazzolo, zoia_dumonteil_mazzolo}. In reactor physics, for instance, knowledge of $\ell_V$ allows assessing the neutron flux due to the chain reaction and hence the deposited power or the number of radiation-induced structural defects~\cite{pazsit, bell}. In a model of epidemics outbreak, $n_V$ corresponds to the number of infections in a region $V$ as a function of the position of the initial infected person (as long as the number of infected people is small, so that nonlinear effects due to the depletion of the susceptibles can be neglected, and that spatial displacements can be described by a simple random walk~\cite{bailey}). The quantity $n_V$ occurs also in population genetics, where one might be interested in quantifying the number $n_V$ of mutations of a given kind $V$, starting from a single character, as a function of the number of generations (this is closely related to the Ewens' formula for the mutation partition, when mutations are allowed to be recurrent~\cite{bertoin}). The goal of this paper is to characterize the statistical properties of the number of visits $n_V$ and of the travelled lengths $\ell_V$ for branching exponential flights.

This paper is structured as follows. In Sec.~\ref{sec_averages} we first focus on the average quantities $\langle \ell_V \rangle$ and $\langle n_V \rangle$. Then, in Sec.~\ref{sec_length} and~\ref{sec_visits} we assess the full distribution of $\ell_V$ and $n_V$, respectively, by resorting to the Feynman-Kac formalism. In particular, we show that recursive formulas for the higher moments of the travelled lengths and for the number of visits can be easily derived based on this approach. Some other significant physical observables associated to $\ell_V $ and $n_V$ are discussed in Sec.~\ref{other_observables}. Then, in Sec.~\ref{sec_rod_model} we illustrate the proposed formalism on one-dimensional exponential flights, the so-called rod model, and support our findings with Monte Carlo simulations. Perspectives are finally presented in Sec.~\ref{conclusions}. Technical details are left to~\ref{fk_length},~\ref{fk_visits} and~\ref{rod_equations}.

\section{The average observables}
\label{sec_averages}

In the following, we introduce a few simplifying hypotheses, whose main advantage is to keep notation to a minimum, yet retaining the key physical mechanisms. Thus, we assume that displacements are performed at a constant speed $v=v_0$, i.e., that only the walker directions $\boldsymbol{\omega}$ do change after collisions. We furthermore assume that the medium is spatially homogeneous, so that $p_k({\mathbf r},v)$ and $\sigma({\mathbf r},v)$ can be taken to be constant. Finally, we assume that the probability density of the directions for the outgoing particles is isotropic, independent of the number of emitted descendants, namely,
\begin{equation}
C_k({\mathbf v}' \to {\mathbf v} |{\mathbf r})=C(\boldsymbol{\omega}' \to \boldsymbol{\omega} |{\mathbf r})=\frac{1}{\Omega_d},
\label{coll_kernel}
\end{equation}
where the normalization factor $\Omega_d=2\pi^{d/2}/\Gamma(d/2)$ is the surface of the unit sphere in dimension $d$.

\begin{figure}[t]
\centerline{ \epsfclipon \epsfxsize=9.0cm
\epsfbox{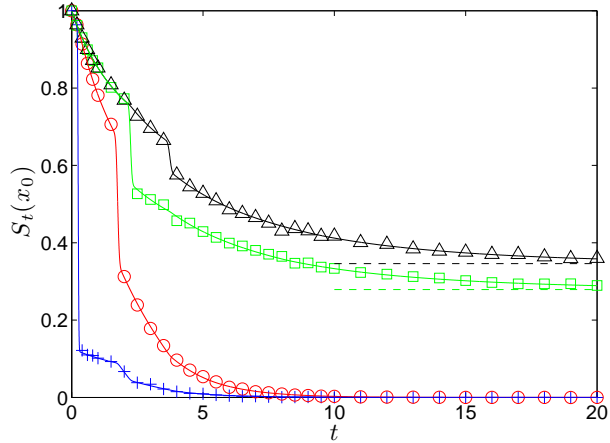} }
\caption{Survival probabilities $S^\pm_t(x_0)$ for $p_0=0.2$, $p_1=0.3$, and $p_2=0.5$ ($\nu_1=1.3$, and $L_c=3.9987$). Blue crosses and red circles: $S^+_t(x_0)$ and $S^-_t(x_0)$, respectively, with $x_0=1.75$ and $L=2$ ($\lambda<0$). Green squares and black triangles: $S^+_t(x_0)$ and $S^-_t(x_0)$, respectively, with $x_0=3.75$ and $L=6$ ($\lambda>0$). Solid lines are numerical integrals of Eq.~\ref{extinction_time_eq}, symbols Monte Carlo simulations with $10^6$ histories. Dashed curves: asymptotic survival probabilities $S_\infty$ from Eq.~\ref{extinction_prob}.}
\label{fig2}
\end{figure}

Branching exponential flights, as defined above, are a Markovian stochastic process that can be observed both as a function of time $\tau$ and discrete generations $n$ (this latter case corresponds to recording the particle position and direction at collision events only). Markovianity is granted by the fact that displacements between collisions are exponentially distributed~\cite{spanier, lux, pazsit}, and implies that knowledge of the phase space variables $\mathbf{r},\boldsymbol{\omega}$ at time $\tau$ or generation $n$ is sufficient to determine the future evolution of the walker~\footnote{Conversely, knowledge of the position $\mathbf{r}$ alone does not ensure Markovianity, as would instead be the case for branching Brownian motion~\cite{derrida, derrida_barrier, spohn, vansaarloos}.}.

To begin with, we address first the average physical observables of branching exponential flights, which in most cases provide a reasonable first-order estimate of the system evolution.

\begin{figure}[t]
\centerline{ \epsfclipon \epsfxsize=9.0cm
\epsfbox{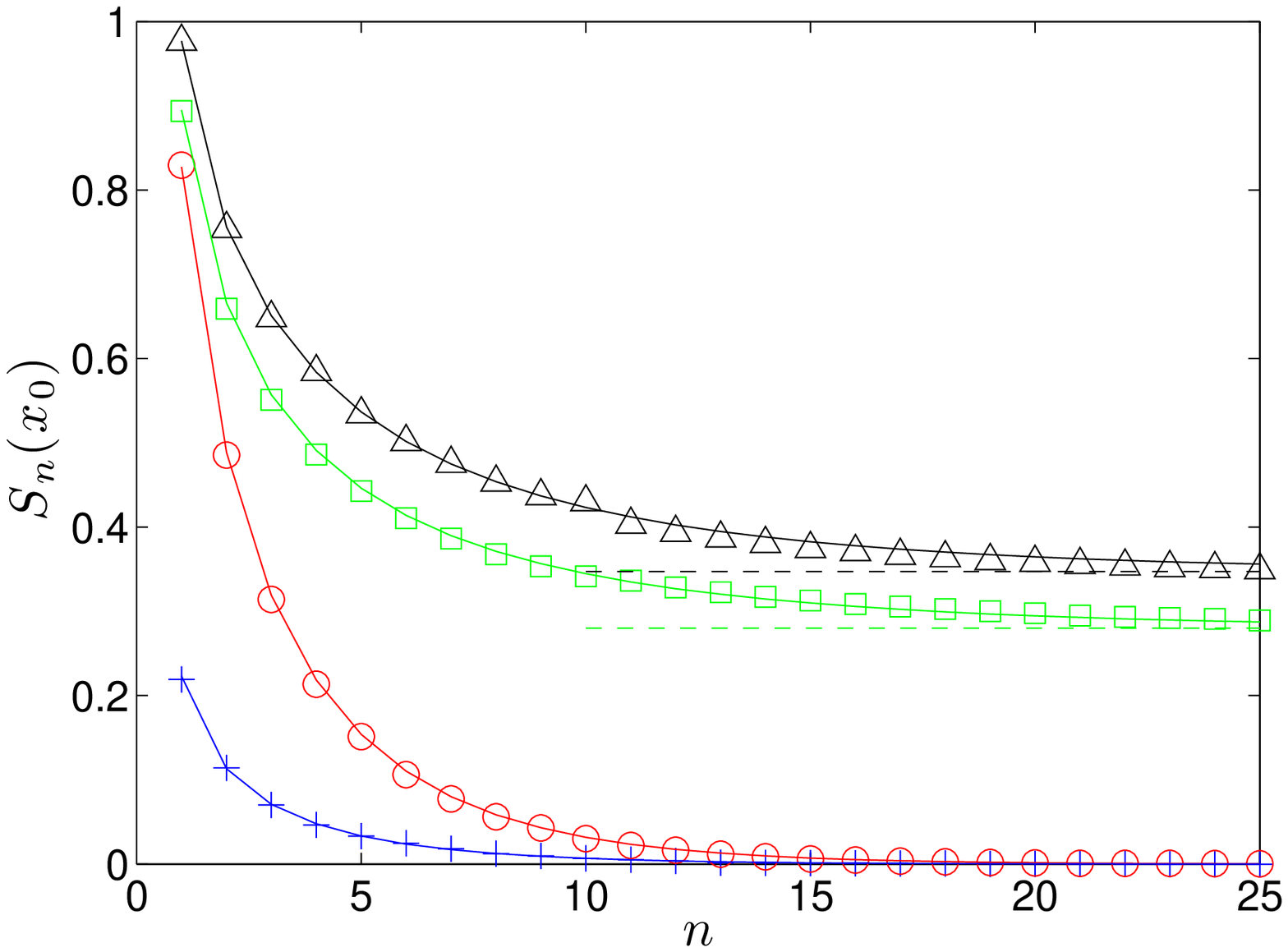} }
\caption{Survival probabilities $S^\pm_n(x_0)$ for $p_0=0.2$, $p_1=0.3$, and $p_2=0.5$ ($\nu_1=1.3$, and $L_c=3.9987$). Blue crosses and red circles: $S^+_n(x_0)$ and $S^-_n(x_0)$, respectively, with $x_0=1.75$ and $L=2$ ($\lambda<0$). Green squares and black triangles: $S^+_n(x_0)$ and $S^-_n(x_0)$, respectively, with $x_0=3.75$ and $L=6$ ($\lambda>0$). Solid lines are numerical integrals of Eq.~\ref{extinction_gen_eq}, symbols Monte Carlo simulations with $10^6$ histories. Dashed curves: asymptotic survival probabilities $S_\infty$ from Eq.~\ref{extinction_prob}.}
\label{fig3}
\end{figure}

\subsection{The average total travelled length $\langle \ell_V \rangle$}

Let $N(\mathbf{r},\boldsymbol{\omega},\tau|\mathbf{r}_0,\boldsymbol{\omega}_0)$ be the average number of particles that at time $\tau$ are found in the phase space element $d\mathbf{r}d\boldsymbol{\omega}$ around $\mathbf{r},\boldsymbol{\omega}$, starting with a single particle emitted at $\mathbf{r}_0$ in direction $\boldsymbol{\omega}_0$, i.e., $N(\mathbf{r},\boldsymbol{\omega},0|\mathbf{r}_0,\boldsymbol{\omega}_0)= q =\delta(\mathbf{r}-\mathbf{r}_0)\delta(\boldsymbol{\omega}-\boldsymbol{\omega}_0)$. Consider a displacement in a small time $d\tau$ along a line oriented as $\boldsymbol{\omega}$: the time variation of $N(\mathbf{r},\boldsymbol{\omega},\tau|\mathbf{r}_0,\boldsymbol{\omega}_0)$ reads
\begin{equation}
\frac{d}{d\tau}N=-v\sigma N+\nu_1\int d\boldsymbol{\omega}' v\sigma C(\boldsymbol{\omega}' \to \boldsymbol{\omega} |{\mathbf r}) N(\mathbf{r},\boldsymbol{\omega}',\tau|\mathbf{r}_0,\boldsymbol{\omega}_0),
\end{equation}
the quantity $\nu_1=\sum_k k p_k$ being the average number of secondary particles emitted per collision event. This equation can be understood as a mass balance: in $d\tau$, $N$ decreases because of particles that at a time rate $v\sigma$ interact, and thus change direction, and increases because of particles that, travelling in another direction $\boldsymbol{\omega}'$, have a collision, are multiplied by a factor $\nu_1$, and change their direction to $\boldsymbol{\omega}$~\cite{bell}. By using $\frac{d}{d\tau}=\frac{\partial}{\partial \tau} + v \boldsymbol{\omega} \cdot \nabla_{\mathbf{r}} $, we finally have
\begin{equation}
\frac{\partial}{\partial \tau}N + v\boldsymbol{\omega} \cdot \nabla_{\mathbf{r}}N=-v\sigma N+\nu_1\int d\boldsymbol{\omega}' v\sigma C(\boldsymbol{\omega}' \to \boldsymbol{\omega} |{\mathbf r}) N(\mathbf{r},\boldsymbol{\omega}',\tau|\mathbf{r}_0,\boldsymbol{\omega}_0),
\end{equation}
which is a Boltzmann-like conservation equation for the average particle density in phase space. Boundary conditions on $N(\mathbf{r},\boldsymbol{\omega},\tau|\mathbf{r}_0,\boldsymbol{\omega}_0)$ depend on the specific problem under analysis. Actually, instead of $N$ it is often common to introduce the quantity $\phi=Nv$, which takes the name of particle flux~\cite{bell}. The stationary behavior of the particle density is provided by integrating over time, and for the stationary flux $\phi(\mathbf{r},\boldsymbol{\omega}|\mathbf{r}_0,\boldsymbol{\omega}_0)=\int_0^\infty d\tau \phi(\mathbf{r},\boldsymbol{\omega},\tau|\mathbf{r}_0,\boldsymbol{\omega}_0)$ we get in particular
\begin{equation}
\boldsymbol{\omega} \cdot \nabla_{\mathbf{r}}\phi + \sigma \phi = \nu_1\int d\boldsymbol{\omega}' \sigma C(\boldsymbol{\omega}' \to \boldsymbol{\omega} |{\mathbf r}) \phi(\mathbf{r},\boldsymbol{\omega}'|\mathbf{r}_0,\boldsymbol{\omega}_0) + q.
\label{eq_phi}
\end{equation}
Eq.~\ref{eq_phi} can be recast in the more compact formula
\begin{equation}
{\cal L}\phi(\mathbf{r},\boldsymbol{\omega}|\mathbf{r}_0,\boldsymbol{\omega}_0) = - q,
\label{op_eq_phi}
\end{equation}
where ${\cal L} = -\boldsymbol{\omega} \cdot \nabla_{\mathbf{r}}- \sigma+  \nu_1\int d\boldsymbol{\omega}' \sigma C(\boldsymbol{\omega}' \to \boldsymbol{\omega} |{\mathbf r}) $ takes the name of (forward) transport operator~\cite{spanier, bell}. The quantity $\phi(\mathbf{r},\boldsymbol{\omega}|\mathbf{r}_0,\boldsymbol{\omega}_0)$ can be intepreted as the stationary density of the total length travelled by the particles in the phase space element $d\mathbf{r}d\boldsymbol{\omega}$ around $\mathbf{r},\boldsymbol{\omega}$: hence, the average travelled length in a given volume $V$ of phase space will be given by
\begin{equation}
\langle \ell_V \rangle (\mathbf{r}_0,\boldsymbol{\omega}_0)= \int d\mathbf{r} \int d\boldsymbol{\omega} V({\mathbf r},\boldsymbol{\omega}) \phi(\mathbf{r},\boldsymbol{\omega}|\mathbf{r}_0,\boldsymbol{\omega}_0),
\end{equation}
where $V({\mathbf r},\boldsymbol{\omega})$ denotes the marker function of the phase space volume $V$, i.e., $V({\mathbf r},\boldsymbol{\omega})=1$ when $\mathbf{r},\boldsymbol{\omega}$ belong to $V$, and $V({\mathbf r},\boldsymbol{\omega})=0$ elsewhere.

\begin{figure}[t]
\centerline{ \epsfclipon \epsfxsize=9.0cm
\epsfbox{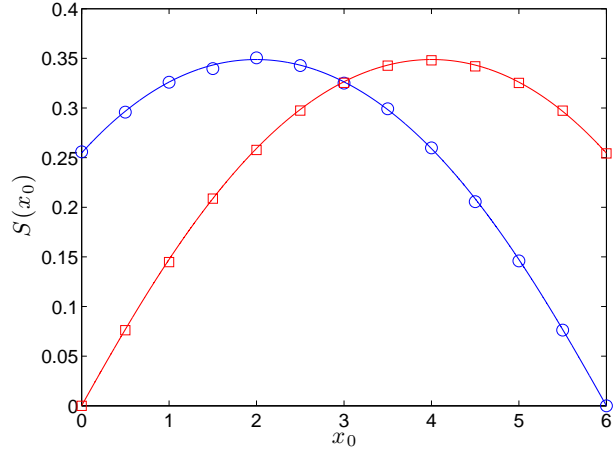} }
\caption{Asymptotic survival probabilities $S^\pm(x_0)$ for $p_0=0.2$, $p_1=0.3$, and $p_2=0.5$ ($\nu_1=1.3$, and $L_c=3.9987$). Blue circles and red squares: $S^+(x_0)$ and $S^-(x_0)$, respectively,with $L=6$ ($\lambda>0$). Solid lines are numerical integrals from Eq.~\ref{extinction_prob}, symbols Monte Carlo simulations with $10^6$ histories.}
\label{fig4}
\end{figure}

\subsection{The average total number of visits $\langle n_V \rangle$}

If generations are considered instead of time, a different mass balance equation for exponential flights can be established. Let $\psi_n(\mathbf{r}, \boldsymbol{\omega}|\mathbf{r}_0, \boldsymbol{\omega}_0)$ be the average number of particles that enter a collision at $\mathbf{r}$, having direction $\boldsymbol{\omega}$, at the $n$-th generation. Then, the following recursive formula can be established
\begin{equation}
\psi_{n+1}=\nu_1 \int d{\mathbf r}'  \int d\boldsymbol{\omega}' T({\mathbf r}' \to {\mathbf r}|\boldsymbol{\omega})C(\boldsymbol{\omega}' \to \boldsymbol{\omega}|{\mathbf r}') \psi_n(\mathbf{r}',\boldsymbol{\omega}'|\mathbf{r}_0,\boldsymbol{\omega}_0),
\end{equation}
with the initial condition $\psi_1(\mathbf{r},\boldsymbol{\omega}|\mathbf{r}_0,\boldsymbol{\omega}_0)=T({\mathbf r}_0 \to {\mathbf r}|\boldsymbol{\omega}_0)$~\cite{bell}. The term $\psi_1$ represents the average particle number entering a collision at the first generation (the so-called uncollided density). The stationary behavior of $\psi_n(\mathbf{r},\boldsymbol{\omega}|\mathbf{r}_0,\boldsymbol{\omega}_0)$ is obtained by summing over all generations: as customary, we define the collision density as being $\psi(\mathbf{r},\boldsymbol{\omega}|\mathbf{r}_0,\boldsymbol{\omega}_0) = \sum_{n=1}^{\infty} \psi_n(\mathbf{r},\boldsymbol{\omega}|\mathbf{r}_0,\boldsymbol{\omega}_0)$~\cite{bell}, and we thus get the integral equation
\begin{equation}
\psi=\nu_1 \int d{\mathbf r}'  \int d\boldsymbol{\omega}' T({\mathbf r}' \to {\mathbf r}|\boldsymbol{\omega})C(\boldsymbol{\omega}' \to \boldsymbol{\omega}|{\mathbf r}') \psi(\mathbf{r}',\boldsymbol{\omega}'|\mathbf{r}_0,\boldsymbol{\omega}_0) + \psi_1 .
\label{eq_psi}
\end{equation}
The quantity $\psi(\mathbf{r},\boldsymbol{\omega}|\mathbf{r}_0,\boldsymbol{\omega}_0)$ physically represents the stationary density of the number of particles entering a collision at $\mathbf{r},\boldsymbol{\omega}$: then, the average number of visits to a given volume $V$ of phase space will be given by
\begin{equation}
\langle n_V \rangle (\mathbf{r}_0,\boldsymbol{\omega}_0) = \int d\mathbf{r} \int d\boldsymbol{\omega} V({\mathbf r},\boldsymbol{\omega}) \psi(\mathbf{r},\boldsymbol{\omega}|\mathbf{r}_0,\boldsymbol{\omega}_0).
\end{equation}
From $\langle n_V \rangle (\mathbf{r}_0,\boldsymbol{\omega}_0)$ and $\langle \ell_V \rangle (\mathbf{r}_0,\boldsymbol{\omega}_0)$ being two average observables of the same stochastic process, and thus closely related to each other, one can easily imagine that the quantities $\psi$ and $\phi$ must be intimately connected as well. Actually, it can be shown that  $\psi(\mathbf{r},\boldsymbol{\omega}|\mathbf{r}_0,\boldsymbol{\omega}_0) =\sigma \phi(\mathbf{r},\boldsymbol{\omega}|\mathbf{r}_0,\boldsymbol{\omega}_0) $~\cite{spanier, bell}: this provides the relation between the stationary densities $N$, $\psi$ and $\phi$, and implies in particular that Eq.~\ref{eq_psi} can be equivalently recast into Eq.~\ref{eq_phi}, by setting $\psi=\sigma \phi$. As a consequence, we have also $\langle n_V \rangle (\mathbf{r}_0,\boldsymbol{\omega}_0) = \int d\mathbf{r} \int d\boldsymbol{\omega} V({\mathbf r},\boldsymbol{\omega}) \sigma \phi(\mathbf{r},\boldsymbol{\omega}|\mathbf{r}_0,\boldsymbol{\omega}_0)$.

\begin{figure}[t]
\centerline{ \epsfclipon \epsfxsize=9.0cm
\epsfbox{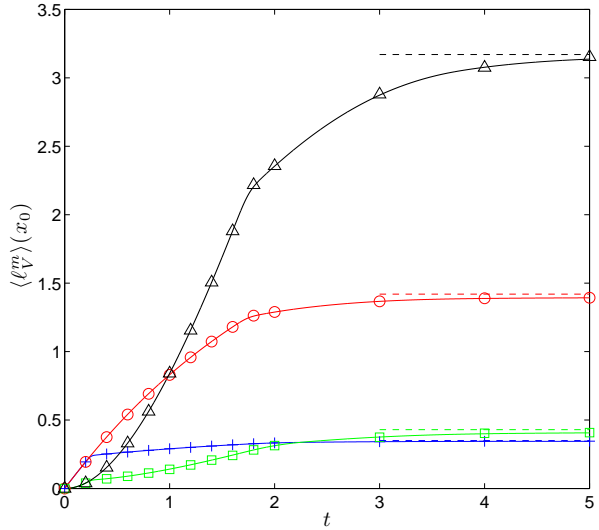} }
\caption{Average and second moment of $\ell_V$, for $x_0=1.75$ and $L=2$, with $p_0=0.2$, $p_1=0.3$, $p_2=0.5$ ($\nu_1=1.3$ and $L<L_c$). Blue crosses: $\langle \ell_V^1 \rangle_t^+(x_0)$; red circles: $\langle \ell_V^1 \rangle_t^-(x_0)$. Green squares: $\langle \ell_V^2 \rangle_t^+(x_0)$; black triangles: $\langle \ell_V^2 \rangle_t^-(x_0)$. Solid lines are numerical integrals from Eq.~\ref{moment_kac_ell_L}, symbols Monte Carlo simulations with $10^6$ histories. Dashed lines are the asymptotic limits in Eq.~\ref{moment_kac_phi}.}
\label{fig5}
\end{figure}

The approach proposed in this Section so to assess the behavior of the average quantities $\langle \ell_V \rangle$ and $\langle n_V \rangle$ can not be straightforwardly extended to higher moments (which are often necessary to quantify the statistical fluctuations around the average), nor to other observables. In the following, we show that this difficulty can be overcome by resorting to the Feynman-Kac formalism, which allows characterizing the full distribution of the variables $\ell_V$ and $n_V$.

\begin{figure}[t]
\centerline{ \epsfclipon \epsfxsize=9.0cm
\epsfbox{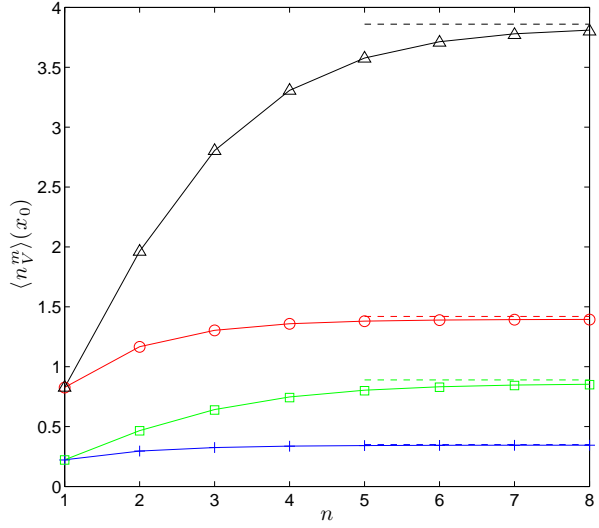} }
\caption{Average and second moment of $n_V$, for $x_0=1.75$ and $L=2$, with $p_0=0.2$, $p_1=0.3$, $p_2=0.5$ ($\nu_1=1.3$ and $L<L_c$). Blue crosses: $\langle n_V^1 \rangle_n^+(x_0)$; red circles: $\langle n_V^1 \rangle_n^-(x_0)$. Green squares: $\langle n_V^2 \rangle_n^+(x_0)$; black triangles: $\langle n_V^2 \rangle_n^-(x_0)$. Solid lines are numerical integrals from Eq.~\ref{moment_eq_n}, symbols Monte Carlo simulations with $10^6$ histories. Dashed lines are the asymptotic limits in Eq.~\ref{moment_kac_phi_n}.}
\label{fig6}
\end{figure}

\section{Total travelled length in $V$}
\label{sec_length}

We formally define the total length $\ell_V(t)$ travelled by a branching exponential flight in a given volume $V$ of the phase space, when observed up to a time $t$, as 
\begin{equation}
\ell_V(t) = \int_0^t V({\mathbf r}',\boldsymbol{\omega}' ) v dt',
\end{equation}
where the integral is intended over all the branching paths of a single realization up to time $t$. The quantity $\ell_V(t)$ is clearly a stochastic variable, which depends on the realizations of the underlying process, as well as on the initial conditions. Instead of studying the probability density function $P_t(\ell_V|{\mathbf r}_0,\boldsymbol{\omega}_0 )$, it is more convenient to introduce the associated moment generating function
\begin{equation}
Q_t(s|{\mathbf r}_0,\boldsymbol{\omega}_0 ) = \langle e^{-s \ell_V(t)}\rangle({\mathbf r}_0,\boldsymbol{\omega}_0),
\end{equation}
where $s$ is the transformed variable with respect to $\ell_V$. Basically, $Q_t(s|{\mathbf r}_0,\boldsymbol{\omega}_0 ) $ can be interpreted as the Laplace transform of $P_t(\ell_V|{\mathbf r}_0,\boldsymbol{\omega}_0 )$. We derive then a backward equation for $Q_t(s|{\mathbf r}_0,\boldsymbol{\omega}_0 ) $ by closely following the approach originally proposed by Kac for Brownian motion~\cite{kac_original}, based on Feynman path integrals~\footnote{The Feynman-Kac formalism more generally applies to continuous-time Markov processes (see, e.g.,~\cite{kac_berkeley, kac_darling, kac, majumdar_review}), and has recently been extended to non-Markovian walks~\cite{barkai, turgeman, barkai_jsp, barkai_non_markov}.}. As detailed in~\ref{fk_length}, the resulting backward Feynman-Kac equation relates the generating function $Q_t$ of the travelled length $\ell_V$ to the generating function $G[z]=\sum_k p_k z^k$ of the offspring number $k$, and reads
\begin{equation}
\frac{1}{v}\frac{\partial}{\partial t}Q_t=\boldsymbol{\omega}_0 \cdot \nabla_{{\mathbf r}_0}Q_t-\sigma Q_t - s V({\mathbf r}_0,\boldsymbol{\omega}_0 ) Q_t + \sigma G[C^*\lbrace Q_t\rbrace],
\label{eq_kac_ell}
\end{equation}
where  $C^*\lbrace Q_t \rbrace$ is a shorthand for the direction-averaged $Q_t$
\begin{equation}
C^*\lbrace Q_t\rbrace = \int C^*(\boldsymbol{\omega}'_0 \to \boldsymbol{\omega}_0 |{\mathbf r}_0) Q_t(s|{\mathbf r}_0,\boldsymbol{\omega}'_0) d\boldsymbol{\omega}'_0,
\end{equation}
$C^*$ being the adjoint probability density with respect to $C$. Equation~\ref{eq_kac_ell} is completed by the initial condition $Q_0(s|{\mathbf r}_0,\boldsymbol{\omega}_0 )=1$ and by the appropriate boundary conditions, which depend on the problem at hand.

\subsection{Moment equations}

Equation~\ref{eq_kac_ell} is a partial differential equation with a nonlinear integral term, for which explicit solutions are hardly available. Moreover, one would still need to invert the solution $Q_t$ so to obtain the probability density of $\ell_V$ in the direct space. A somewhat simpler approach consists in deriving the corresponding moment equations\footnote{For Brownian motion without branching, a similar approach was proposed by Kac~\cite{kac} and later extended, e.g., in~\cite{berezhkovskii, agmon_lett, agmon}. Similarly, the moments of exponential flights without branching are discussed, e.g., in~\cite{blanco, mazzolo, benichou_epl, zoia_dumonteil_mazzolo, zdm_pre}.}: by the definition of $Q_t$, the moments of the travelled length can be obtained from
\begin{equation}
\langle \ell_V^m \rangle_t ({\mathbf r}_0,\boldsymbol{\omega}_0) = (-1)^m \frac{\partial^m}{\partial s^m} Q_t(s|{\mathbf r}_0,\boldsymbol{\omega}_0)\vert_{s=0}.
\end{equation}
By taking the $m$-th derivative of Eq.~\ref{eq_kac_ell} and resorting to the Fa\`{a} di Bruno's formula for multiple derivatives of composite functions~\cite{pitman_book}, we get the following recursive formula for the moments of the trace length
\begin{equation}
\frac{1}{v}\frac{\partial}{\partial t}\langle \ell_V^m \rangle_t ={\cal L}^*\langle \ell_V^m \rangle_t + m V({\mathbf r}_0,\boldsymbol{\omega}_0 ) \langle \ell_V^{m-1} \rangle_t  +  \sigma \sum_{j=2}^m \nu_j {\cal B}_{m,j} \left[ C^*\lbrace \langle \ell_V^i \rangle_t \rbrace \right],
\label{moment_kac_ell_L}
\end{equation}
for $m \ge 1$, where
\begin{equation}
{\cal L}^* = \boldsymbol{\omega}_0 \cdot \nabla_{{\mathbf r}_0} - \sigma + \sigma \nu_1 \int C^*(\boldsymbol{\omega}'_0 \to \boldsymbol{\omega}_0 |{\mathbf r}_0) d\boldsymbol{\omega}'_0
\end{equation}
is the (backward) transport operator adjoint to ${\cal L}$~\cite{bell}. Here ${\cal B}_{m,j} \left[z_{i}\right]={\cal B}_{m,j} \left[z_{1},z_{2},\cdots,z_{m-j+1}\right]$ are the Bell's polynomials~\cite{pitman_book}, and $\nu_j=\langle k(k-1)...(k-j+1)\rangle$ are the falling factorial moments of the descendant number, with $\nu_0=1$. Bell polynomials~\footnote{The first few polynomials read: ${\cal B}_{0,0}=1$; ${\cal B}_{1,1} [z_1]=z_1; {\cal B}_{2,1} [z_1,z_2]=z_2, {\cal B}_{2,2} [z_1,z_2]=z^2_1; {\cal B}_{3,1} [z_1,z_2,z_3]=z_3, {\cal B}_{3,2} [z_1,z_2,z_3]=3z_1z_2, {\cal B}_{3,3}[z_1,z_2,z_3]=z_1^3; ...$.} commonly appear in connection with the combinatorics of branched structures~\cite{pitman_book}: this might give a hint about their role in Eq.~\ref{moment_kac_ell_L}, which relates the moments $\langle \ell_V^{m} \rangle_{t}$ of the travelled length to the moments $\nu_j$ of the descendant number. The recurrence is initiated with the conditions $\langle \ell_V^{0} \rangle_t=1$ (from normalization), and $\langle \ell_V^{m} \rangle_{0}=0$. Observe that $\langle \ell_V^{1} \rangle_{t}$ depends only on $\nu_1$, $\langle \ell_V^{2} \rangle_{t}$ on $\nu_1$ and $\nu_2$, and so on.

\subsection{Stationary behavior}

Most often, the observation time $t$ is much longer than the characteristic time scale of the system dynamics, which means that trajectories are followed up to $t \to \infty$. In this case, the time derivative in Eq.~\ref{moment_kac_ell_L} vanishes, provided that the moment $\langle \ell_V^m \rangle_t$ does not diverge when $t \to \infty$. We therefore get a recursive formula for the stationary moments $\langle \ell_V^m \rangle = \lim_{t \to \infty} \langle \ell_V^m \rangle_t$, namely,
\begin{equation}
{\cal L}^*\langle \ell_V^m \rangle({\mathbf r}_0,\boldsymbol{\omega}_0) = - U_{m-1}({\mathbf r}_0,\boldsymbol{\omega}_0), 
\label{moment_kac_ell_stat}
\end{equation}
where
\begin{equation}
U_{m-1}({\mathbf r}_0,\boldsymbol{\omega}_0) = m V({\mathbf r}_0,\boldsymbol{\omega}_0 ) \langle \ell_V^{m-1} \rangle  + \sigma \sum_{j=2}^m \nu_j {\cal B}_{m,j} \left[ C^*\lbrace \langle \ell_V^i \rangle \rbrace \right]
\end{equation}
can be interpreted as a (known) source term that depends at most on the moments of order $m-1$. Now, from ${\cal L}^*$ being the adjoint operator with respect to ${\cal L}$, if one can solve Eq.~\ref{op_eq_phi} for a point source $q=\delta(\mathbf{r}-\mathbf{r}_0)\delta(\boldsymbol{\omega}-\boldsymbol{\omega}_0)$ and obtain the corresponding stationary flux $\phi$, then Eq.~\ref{moment_kac_ell_stat} can be explicitly inverted, and gives
\begin{equation}
\langle \ell_V^m \rangle ({\mathbf r}_0,\boldsymbol{\omega}_0) = \int d {\mathbf r}\int d\boldsymbol{\omega} \phi({\mathbf r},\boldsymbol{\omega}|{\mathbf r}_0,\boldsymbol{\omega}_0) U_{m-1}({\mathbf r},\boldsymbol{\omega}),
\label{moment_kac_phi}
\end{equation}
which means that the stationary moments of the travelled length can be obtained by convoluting the stationary flux with the source term $U_{m-1}$. As a particular case, when $p_{k \ge 2}=0$ we reobtain the simpler recursive formula derived in~\cite{benichou_epl, zdm_pre} for non-branching exponential flights.

Finally, for the average length travelled in $V$, i.e., $m=1$, we recover the formula $\langle \ell_V^1 \rangle = \int d {\mathbf r}\int d\boldsymbol{\omega} V({\mathbf r},\boldsymbol{\omega}) \phi({\mathbf r},\boldsymbol{\omega}|{\mathbf r}_0,\boldsymbol{\omega}_0)$, since $U_{0}({\mathbf r},\boldsymbol{\omega})=V({\mathbf r},\boldsymbol{\omega})$.

\section{Total number of visits to $V$}
\label{sec_visits}

We address then the statistical properties of the total number of visits $n_V(n)$ performed by a branching exponential flight in a given volume $V$ of the phase space, when observed up to the $n$-th generation\footnote{The analysis of discrete (isotropic) branching walks can be extended to arbitrary flight length densities $T$, as shown in~\cite{zdm_pre_disc_kac, zdm_epl}, by resorting to an integral formulation. Here we stuck however to exponential flights, which imposes $T$ as given in Eq.~\ref{exp_kernel}.}. We formally define
\begin{equation}
n_V(n) = \sum_i V({\mathbf r}_i,\boldsymbol{\omega}_i ),
\end{equation}
where the sum is intended over all the points visited by the branching path up to entering the $n$-th generation. We adopt the convention that the source is not taken into account. The quantity $n_V(n)$ is again a stochastic variable depending on the realizations of the underlying process and on the initial conditions. Similarly as done for $\ell_V$, instead of studying the probability $P_n(n_V|{\mathbf r}_0,\boldsymbol{\omega}_0 )$, it is more convenient to introduce the associated moment generating function
\begin{equation}
Q_n(u|{\mathbf r}_0,\boldsymbol{\omega}_0 ) = \langle e^{-u n_V(n)}\rangle({\mathbf r}_0,\boldsymbol{\omega}_0),
\end{equation}
where $u$ is the transformed variable with respect to $n_V$. As shown in~\ref{fk_visits}, the backward discrete Feynman-Kac equation for $Q_n(u|{\mathbf r}_0,\boldsymbol{\omega}_0 ) $ relates the generating function $Q_n$ of the number of visits $n_V$ to the generating function $G$ of the offspring number $k$, and reads
\begin{equation}
-\boldsymbol{\omega}_0 \cdot \nabla_{{\mathbf r}_0} Q_{n+1}(u|{\mathbf r}_0,\boldsymbol{\omega}_0) + \sigma Q_{n+1}(u|{\mathbf r}_0,\boldsymbol{\omega}_0) = \sigma e^{-u V({\mathbf r}_0,\boldsymbol{\omega}_0)} G\left[ C^*\lbrace Q_n \rbrace \right],
\label{equation_F}
\end{equation}
with the initial condition
\begin{equation}
Q_1(u|{\mathbf r}_0,\boldsymbol{\omega}_0)=\int e^{-u V({\mathbf r}_1,\boldsymbol{\omega}_0)} T^*({\mathbf r}_1 \to {\mathbf r}_0 |\boldsymbol{\omega}_0)d{\mathbf r}_1 ,
\end{equation}
and the appropriate boundary conditions. We have used the shorthand $C^*\lbrace Q_n\rbrace=\int C^*(\boldsymbol{\omega}'_0 \to \boldsymbol{\omega}_0 |{\mathbf r}_1) Q_n(u|{\mathbf r}_1,\boldsymbol{\omega}_0) d\boldsymbol{\omega}'_0$. As a particular case, when particles can not move (i.e., $\sigma \to \infty$), spatial dependences can be neglected and from Eq.~\ref{equation_F} one recovers the counting statistics of a simple Galton-Watson process observed up to the $n$-th generation~\cite{harris}.

\subsection{Moment equations}

Equation~\ref{equation_F} is a nonlinear integro-differential and finite differences equation. Similarly as in the case of $\ell_V$, the analysis of the distribution of $n_V$ can be simplified by deriving the corresponding moment equations: by the definition of $Q_n$, the moments of the number of visits can be obtained from
\begin{equation}
\langle n_V^{m} \rangle_n ({\mathbf r}_0,\boldsymbol{\omega}_0) = (-1)^m \frac{\partial^m}{\partial u^m} Q_n(u|{\mathbf r}_0,\boldsymbol{\omega}_0)\vert_{u=0}.
\end{equation}
Then, by taking the $m$-th derivative of Eq.~\ref{equation_F}, we get the following recursive formula for the moments of number of visits
\begin{eqnarray}
-\boldsymbol{\omega}_0 \cdot \nabla_{{\mathbf r}_0}  \langle n_V^{m} \rangle_{n+1}+ \sigma \langle n_V^{m} \rangle_{n+1} =
\sigma \sum_{j=1}^m \nu_j {\cal B}_{m,j} \left[ C^*\lbrace \langle n_V^{i} \rangle_n \rbrace \right]+\nonumber \\
\sigma \sum_{k=1}^m  {m \choose k} V({\mathbf r}_0,\boldsymbol{\omega}_0) \sum_{j=0}^{m-k} \nu_j {\cal B}_{m-k,j} \left[ C^*\lbrace \langle n_V^{i} \rangle_n \rbrace \right],
\label{moment_eq_n}
\end{eqnarray}
for $m \ge 1$. Equation~\ref{moment_eq_n} relates the moments $\langle n_V^{m} \rangle_{n}$ of the number of visits to the moments $\nu_j$ of the descendant number. Observe that $\langle n_V^{1} \rangle_{n}$ depends only on $\nu_1$, $\langle n_V^{2} \rangle_{n}$ on $\nu_1$ and $\nu_2$, and so on. The recurrence is initiated with the conditions $\langle n_V^{0} \rangle_n=1$ (from normalization), and $\langle n_V^{m} \rangle_{1}=\int V({\mathbf r}_1,\boldsymbol{\omega}_0) T^*({\mathbf r}_1 \to {\mathbf r}_0 |\boldsymbol{\omega}_0)d{\mathbf r}_1 $.

\subsection{Stationary behavior}

Most often, one considers trajectories that are followed up to $n \to \infty$, provided that the moment $\langle n_V^{m} \rangle_n$ does not diverge. We therefore get a recursive formula for the stationary moments $\langle n_V^{m}  \rangle = \lim_{n \to \infty} \langle n_V^{m}  \rangle_n$, namely,
\begin{equation}
{\cal L}^*\langle n_V^{m}  \rangle({\mathbf r}_0,\boldsymbol{\omega}_0) = - H_{m-1}({\mathbf r}_0,\boldsymbol{\omega}_0),
\label{moment_kac_n_stat}
\end{equation}
where 
\begin{equation}
H_{m-1} =\sigma \sum_{k=1}^m  {m \choose k} V({\mathbf r}_0,\boldsymbol{\omega}_0) \sum_{j=0}^{m-k} \nu_j {\cal B}_{m-k,j} \left[ C^*\lbrace \langle n_V^{i} \rangle \rbrace \right]+\sigma \sum_{j=2}^m \nu_j {\cal B}_{m,j} \left[ C^*\lbrace \langle n_V^{i} \rangle \rbrace \right].
\end{equation}
is a source term, and we have singled out the term of order $m$ in the Bell polynomials. The quantity $H_{m-1}$ is closely related to $U_{m-1}$, and the contribution $\sigma \sum_{j=2}^m \nu_j {\cal B}_{m,j} \left[ C^*\lbrace \langle n_V^{i} \rangle \rbrace \right]$ is common to both. When $m=1$, we have $H_0=\sigma U_0$.

As done for the moments of travelled lengths, Eq.~\ref{moment_kac_n_stat} can be explicitly inverted in terms of the corresponding stationary flux $\phi$, and gives
\begin{equation}
\langle n_V^{m} \rangle ({\mathbf r}_0,\boldsymbol{\omega}_0) = \int d {\mathbf r}\int d\boldsymbol{\omega} \phi({\mathbf r},\boldsymbol{\omega}|{\mathbf r}_0,\boldsymbol{\omega}_0) H_{m-1}({\mathbf r},\boldsymbol{\omega}),
\label{moment_kac_phi_n}
\end{equation}
which means that the stationary moments of the number of visits can be obtained by convoluting the stationary flux with the source term $H_{m-1}$.

In particular, for the average number of visits to $V$ ($m=1$) we recover the formula $\langle n_V^{1} \rangle = \int d {\mathbf r}\int d\boldsymbol{\omega} V({\mathbf r}_0,\boldsymbol{\omega}_0) \sigma \phi({\mathbf r},\boldsymbol{\omega}|{\mathbf r}_0,\boldsymbol{\omega}_0)$, since $H_{0}({\mathbf r},\boldsymbol{\omega})=\sigma V({\mathbf r}_0,\boldsymbol{\omega}_0)$. The close relation between $\langle n_V^1 \rangle$ and $\langle \ell_V^1 \rangle$ carries over to moments of any order $m \ge 1$ (via Eq.~\ref{moment_kac_n_stat}), although the simple proportionality that holds true for the average does not apply to higher moments.

\section{Other physical observables}
\label{other_observables}

The Feynman-Kac approach proposed in the previous Sections allows fully characterizing the distribution of the travelled length and of the number of visits. In the following, we show that a number of other interesting features of the process can be assessed by relying upon the same formalism, with minimal modifications, and we discuss some significant examples.

\subsection{Probability of never visiting a region $V$}

For instance, one might be interested in determining the probability $R_n({\mathbf r}_0,\boldsymbol{\omega}_0)$ that a branching exponential flight coming from a point source at ${\mathbf r}_0,\boldsymbol{\omega}_0$ never collides in a given domain $V$, up to generation $n$. This is intimately related to the well-known gambler's ruin problem~\cite{hughes, weiss}. It follows that $R_n({\mathbf r}_0,\boldsymbol{\omega}_0) = P_n(n_V=0|{\mathbf r}_0,\boldsymbol{\omega}_0 )$. Now, if we introduce the probability generating function $\langle u^{n_V(n)}\rangle({\mathbf r}_0,\boldsymbol{\omega}_0) = \sum_i u^i P_n(n_V=i|{\mathbf r}_0,\boldsymbol{\omega}_0 )$, then by construction $R_n({\mathbf r}_0,\boldsymbol{\omega}_0)=\langle u^{n_V(n)}\rangle({\mathbf r}_0,\boldsymbol{\omega}_0)\vert_{u=0}$. By comparing $\langle u^{n_V(n)}\rangle({\mathbf r}_0,\boldsymbol{\omega}_0) $ to $Q_n(u|{\mathbf r}_0,\boldsymbol{\omega}_0 ) =\langle e^{-u n_V(n)}\rangle({\mathbf r}_0,\boldsymbol{\omega}_0) $, it is apparent that $ R_n({\mathbf r}_0,\boldsymbol{\omega}_0) $ satisfies
\begin{equation}
-\boldsymbol{\omega}_0 \cdot \nabla_{{\mathbf r}_0}R_{n+1}({\mathbf r}_0,\boldsymbol{\omega}_0)  + \sigma R_{n+1}({\mathbf r}_0,\boldsymbol{\omega}_0) = \sigma \bar{V}({\mathbf r}_0,\boldsymbol{\omega}_0) G\left[ C^*\left\lbrace R_n \right\rbrace \right],
\end{equation}
where $\bar{V}({\mathbf r}_0,\boldsymbol{\omega}_0)=1-V({\mathbf r}_0,\boldsymbol{\omega}_0)$. As for the initial conditions, we have $R_1({\mathbf r}_0,\boldsymbol{\omega}_0)=\int \bar{V}({\mathbf r}_1,\boldsymbol{\omega}_0) T^*({\mathbf r}_1 \to {\mathbf r}_0 |\boldsymbol{\omega}_0)d{\mathbf r}_1$. When $n \to \infty$, we get the stationary probability equation
\begin{equation}
-\boldsymbol{\omega}_0 \cdot \nabla_{{\mathbf r}_0}R({\mathbf r}_0,\boldsymbol{\omega}_0)  + \sigma R({\mathbf r}_0,\boldsymbol{\omega}_0) = \sigma \bar{V}({\mathbf r}_0,\boldsymbol{\omega}_0) G\left[ C^*\left\lbrace R \right\rbrace \right],
\end{equation}
where we have set $R({\mathbf r}_0,\boldsymbol{\omega}_0)=\lim_{n \to \infty} R_n({\mathbf r}_0,\boldsymbol{\omega}_0)$.

\subsection{Survival probability}

The quantities $\ell_V$ and $n_V$, as defined above, are cumulative, i.e., their knowledge integrates the whole history of the walk, from the source to the measure time, or generation. Sometimes, it is necessary to provide information about local (instantaneous) countings, namely the number $m_V(t)$ and $m_V(n)$ of particles in the volume $V$ when an observation is performed at time $t$, or at generation $n$, respectively (see, e.g.,~\cite{pazsit}). As shown in~\ref{fk_length}, the probability generating function $W_t(s|{\mathbf r}_0,\boldsymbol{\omega}_0)= \langle s^{m_V(t)}\rangle$ satisfies
\begin{equation}
\frac{1}{v}\frac{\partial}{\partial t}W_t(s|{\mathbf r}_0,\boldsymbol{\omega}_0 )=\boldsymbol{\omega}_0 \cdot \nabla_{{\mathbf r}_0}W_t(s|{\mathbf r}_0,\boldsymbol{\omega}_0 )-\sigma W_t(s|{\mathbf r}_0,\boldsymbol{\omega}_0 )+ \sigma G[C^*\lbrace W_t \rbrace],
\label{extinction_time}
\end{equation}
with initial condition $W_0(s|{\mathbf r}_0,\boldsymbol{\omega}_0)=s^{V({\mathbf r}_0,\boldsymbol{\omega}_0)}$. Equation~\ref{extinction_time} is known in reactor physics as the P\'{a}l-Bell equation~\cite{pal, bell_nuc, pazsit}. Furthermore, as detailed in~\ref{fk_visits}, the probability generating function $W_n(u|{\mathbf r}_0,\boldsymbol{\omega}_0)= \langle u^{m_V(n)}\rangle$ satisfies
\begin{equation}
-\boldsymbol{\omega}_0 \cdot \nabla_{{\mathbf r}_0} W_{n+1}(u|{\mathbf r}_0,\boldsymbol{\omega}_0) + \sigma W_{n+1}(u|{\mathbf r}_0,\boldsymbol{\omega}_0) = \sigma G\left[ C^*\lbrace W_n \rbrace \right],
\label{extinction_gen}
\end{equation}
with initial condition $W_1(u|{\mathbf r}_0,\boldsymbol{\omega}_0)=\int u^{V({\mathbf r}_1,\boldsymbol{\omega}_0)} T^*({\mathbf r}_1 \to {\mathbf r}_0 |\boldsymbol{\omega}_0)d{\mathbf r}_1$.

Assume now that $V$ is bounded, i.e., particles are lost upon leaving the volume: then, we might want to assess the survival probability at time $t$ or generation $n$, due to the interplay between the branching mechanism and the spatial leakages. As particles can not re-enter $V$ after crossing the boundaries, if $m_V(t)=0$, then also $m_V(t')=0$ for $t' \ge t$ (i.e., the process goes to extinction), and the same holds for $m_V(n)$. Hence, by definition, the probability of having zero particles in the volume $V$ at a time $t$ is given by $W_t(s=0|{\mathbf r}_0,\boldsymbol{\omega}_0)$, which equivalently yields the probability that extinction is reached for times smaller than $t$, since $V$ is bounded~\cite{pazsit}. We define then the survival probability as $S_t({\mathbf r}_0,\boldsymbol{\omega}_0)=1-W_t(s=0|{\mathbf r}_0,\boldsymbol{\omega}_0)$, which by direct substitution in Eq.~\ref{extinction_time} satisfies
\begin{equation}
\frac{1}{v}\frac{\partial}{\partial t}S_t({\mathbf r}_0,\boldsymbol{\omega}_0 )=\boldsymbol{\omega}_0 \cdot \nabla_{{\mathbf r}_0}S_t({\mathbf r}_0,\boldsymbol{\omega}_0 )-\sigma S_t({\mathbf r}_0,\boldsymbol{\omega}_0 )- \sigma F[C^*\lbrace S_t \rbrace],
\label{extinction_time_eq}
\end{equation}
where we set $F[z]=\sum_{k=1}^\infty \alpha_k z^k$, with $\alpha_k=(-1)^k \nu_k / k! $~\cite{bell_nuc}. At the boundaries, $S_t$ must vanish when $\boldsymbol{\omega}_0$ is directed towards the exterior of $V$. The probability of having zero particles in the volume $V$ at generation $n$ is similarly given by $W_n(u=0|{\mathbf r}_0,\boldsymbol{\omega}_0)$, which therefore yields the probability that extinction is reached for generations smaller than $n$. We define then the associated survival probability as $S_n({\mathbf r}_0,\boldsymbol{\omega}_0)=1-W_n(s=0|{\mathbf r}_0,\boldsymbol{\omega}_0)$, which by direct substitution in Eq.~\ref{extinction_gen} satisfies
\begin{equation}
\boldsymbol{\omega}_0 \cdot \nabla_{{\mathbf r}_0} S_{n+1}({\mathbf r}_0,\boldsymbol{\omega}_0) - \sigma S_{n+1}({\mathbf r}_0,\boldsymbol{\omega}_0) = \sigma F\left[ C^*\lbrace S_n \rbrace \right],
\label{extinction_gen_eq}
\end{equation}
where $S_n$ must again vanish at the boundaries when $\boldsymbol{\omega}_0$ is directed towards the exterior of $V$. Finally, by either taking the limit $S=\lim_{t \to \infty} S_t({\mathbf r}_0,\boldsymbol{\omega}_0)$ or $S=\lim_{n \to \infty} S_n({\mathbf r}_0,\boldsymbol{\omega}_0)$, respectively, the probability of ultimate survival $S({\mathbf r}_0,\boldsymbol{\omega}_0)$ satisfies
\begin{equation}
\boldsymbol{\omega}_0 \cdot \nabla_{{\mathbf r}_0}S - \sigma S= \sigma F[C^*\lbrace S \rbrace].
\label{extinction_prob}
\end{equation}
Observe that in principle $S=0$ is always a solution to Eq.~\ref{extinction_prob}, which would imply almost sure extinction, i.e., a vanishing probability that infinitely long branching chains exist in $V$. From the Galton-Watson theory, we know that when $\nu_1 \le 1$ the branching process goes to extinction even in the absence of spatial leakages, hence $S=0$. However, when $\nu_1 >1$ the branching process would grow indefinitely, and it may happen that the particle loss due to finite geometry is not sufficient to compensate the population growth. In this case, the solution $S=0$ would become unstable, and $S_t$ (or $S_n$) would converge towards a nontrivial survival probability $S = S_{\infty} >0$. The stability analysis of the solution $S=0$ can be carried out by introducing a small perturbation, for instance in the form $\hat{S} \simeq \epsilon X(t) Y({\mathbf r}_0,\boldsymbol{\omega}_0)$, the amplitude $\epsilon >0$ being a small positive constant, with $X(t)>0$ and $Y({\mathbf r}_0,\boldsymbol{\omega}_0)>0$. Now, if we inject $\hat{S}$ into Eq.~\ref{extinction_time_eq}, and take the limit $\epsilon \to 0$, we obtain an equation for the perturbation amplitude
\begin{equation}
\frac{1}{v}\frac{1}{X(t)}\frac{\partial X(t)}{\partial t}=\frac{\boldsymbol{\omega}_0 \cdot \nabla_{{\mathbf r}_0}Y({\mathbf r}_0,\boldsymbol{\omega}_0 )-\sigma Y({\mathbf r}_0,\boldsymbol{\omega}_0 )+ \sigma \nu_1 C^*\lbrace Y \rbrace}{Y},
\label{stability}
\end{equation}
where at the numerator of the right hand side we recognize the adjoint operator ${\cal L}^*$. From the separation of the variables, Eq.~\ref{stability} shows that the evolution of the perturbation amplitude with respect to time is determined by the ratio $\lambda = {\cal L}^*Y/Y$, hence by the eigenvalue equation ${\cal L}^*Y = \lambda Y$. The spectrum of the eigenvalues of ${\cal L}^*$ depends on the geometry of $V$ and on the boundary conditions. If all eigenvalues $\lambda$ are negative, the amplitude of the small perturbation will shrink in time, so that eventually $S=0$; if on the contrary at least one eigenvalue is positive, then the small perturbation will grow in time, which means that $S=0$ is unstable, and eventually $S=S_{\infty}$. For a given branching process, the crossover between these two regimes depends on the size and shape of the volume $V$, and generally speaking one would expect that $S=S_{\infty}$ for a volume size larger than some critical value $V_c$~\cite{bell, pazsit}, which is attained when $\lambda=0$. When $\nu_1 \le 1$, $V_c = \infty$. The stability analysis of $\hat{S} \simeq \epsilon X_n Y({\mathbf r}_0,\boldsymbol{\omega}_0)$ in Eq.~\ref{extinction_gen_eq} leads to
\begin{equation}
\frac{X_{n+1}}{X_n}=-\frac{\sigma \nu_1 C^*\lbrace Y \rbrace}{\boldsymbol{\omega}_0 \cdot \nabla_{{\mathbf r}_0}Y({\mathbf r}_0,\boldsymbol{\omega}_0 )-\sigma Y({\mathbf r}_0,\boldsymbol{\omega}_0 )}.
\label{stability_gen}
\end{equation}
The evolution of the perturbation amplitude is determined then by the eigenvalue equation $\boldsymbol{\omega}_0 \cdot \nabla_{{\mathbf r}_0}Y({\mathbf r}_0,\boldsymbol{\omega}_0 )-\sigma Y({\mathbf r}_0,\boldsymbol{\omega}_0 )+\frac{\sigma \nu_1}{\beta} C^*\lbrace Y \rbrace =0$: when $\beta<1$ the pertubation shrinks and when $\beta>1$ the pertubation grows, the crossover occurring for a critical volume $V_c$ such that $\beta=1$.

\section{The rod model}
\label{sec_rod_model}

In order to illustrate the previous results, we revisit here a relevant example inspired by reactor physics. Neutrons in a multiplying medium undergo branching exponential flights, where conceptually radiative capture represents absorption ($p_0$), scattering corresponds to $p_1$ and fission to $p_{k\ge 2}$: in realistic situations, the cross sections and all the other physical parameters depend on energy and position, and angular distributions are often mildy or even strongly anisotropic~\cite{bell}. To simplify the matter, we assume that cross sections are constant, scattering is isotropic and particles travel at an average constant speed $v=1$. Moreover, we address a one-dimensional configuration, namely the interval $[0,L]$, where only two angular directions (forward or backward) are allowed: all physical quantities will be then denoted with a superscript $\pm$ according to whether $\omega$ is taken along the $x$-axis ($+$), or in the opposite direction ($-$). Despite these many simplifications, the so-called rod model yet captures the key features of neutron transport, and has been widely adopted in neutronics~\cite{harris, wing}.

To begin with, we impose leakage boundary conditions at $x=0$ and $x=L$ and study the survival probabilities. By setting $Y^+(x_0 )=Y(x_0,\omega_0=+)$ and $Y^-(x_0 )=Y(x_0,\omega_0=-)$, the eigenvalue equation ${\cal L}^*Y = \lambda Y$ gives rise to a system of two coupled linear differential equations
\begin{eqnarray}
\frac{\partial}{\partial x_0}Y^+(x_0)-\sigma Y^+(x_0 )+ \frac{\sigma \nu_1 }{2}\left(Y^+(x_0 ) + Y^-(x_0 ) \right)=\lambda Y^+(x_0 ),\nonumber \\
-\frac{\partial}{\partial x_0}Y^-(x_0)-\sigma Y^-(x_0 )+ \frac{\sigma \nu_1 }{2}\left(Y^+(x_0 ) + Y^-(x_0 ) \right)= \lambda Y^-(x_0 ),
\label{system_y}
\end{eqnarray}
where we have used $C^*\lbrace Y \rbrace=(Y^+(x_0 ) + Y^-(x_0 ))/2$, thanks to isotropy. The general integrals $Y^+(x_0 )=Y_1(x_0;\kappa_1,\kappa_2)$ and $Y^-(x_0 )=Y_2(x_0;\kappa_1,\kappa_2)$ of Eq.~\ref{system_y} can be easily obtained as a combination of exponential functions, up to two integration constants, say $\kappa_1$ and $\kappa_2$, that are to be imposed by boundary conditions. Due to the leakages, at the boundaries we have $Y^+(L)=0$ and $Y^-(0)=0$. The solutions $Y^+=Y^-=0$ clearly satisfy the system in Eq.~\ref{system_y}, together with boundary conditions. Searching for nontrivial solutions leads then to solving
\begin{equation}
\det \left( \begin{array}{c}
Y_1(L;\kappa_1,\kappa_2)=0 \\
Y_2(0;\kappa_1,\kappa_2)=0 \\
\end{array} \right)=0
\end{equation}
with respect to the basis of $\kappa_1$ and $\kappa_2$. This in turn gives an implicit equation for the eigenvalues $\lambda$ as a function of the system parameters $L$, $\sigma$, and $\nu_1$, namely,
\begin{eqnarray}
\cosh \left( L\sigma\sqrt{ \lambda_\sigma     \left(\lambda_\sigma-\nu_1 \right)} \right) +\frac{\left(\lambda_\sigma-\frac{\nu_1}{2} \right)\sinh \left( L\sigma\sqrt{\lambda_\sigma \left(\lambda_\sigma-\nu_1 \right)} \right)  }{\sqrt{\lambda_\sigma\left(\lambda_\sigma-\nu_1 \right)} }=0,
\label{eq_lambda}
\end{eqnarray}
where we have set the dimensionless variable $\lambda_\sigma=\lambda/\sigma+1$.

When $\lambda=\lambda(L,\sigma,\nu_1)<0$, then $S=0$ and the neutron chain reaction will die out because of leakages and possibly absorptions; when on the contrary $\lambda>0$, then $S=S_{\infty}(x_0)$, and the chain reaction will diverge, as particles born from fission are not sufficiently compensated by leakages and absorptions; the crossover between these two regimes is reached for $\lambda=0$, which therefore defines the portion of the parameter space ($L$, $\sigma$, $\nu_1$) for which the branching process will attain ultimate extinction. When $\nu_1 \le 1$, all $\lambda$ stay negative. When $\nu_1>1$, imposing $\lambda=0$ in Eq.~\ref{eq_lambda} yields the explicit relation
\begin{equation}
L_c=L_{\lambda=0}=\frac{2}{\sigma} \frac{\tan^{-1}\left( \frac{1}{\sqrt{\nu_1-1}}\right)}{\sqrt{\nu_1-1}},
\label{critical_length}
\end{equation}
where $L_c$ is the maximum system size such that for $L > L_c$ there exists a finite survival probability that the neutron population will grow indefinitely. By resorting to the same arguments, the analysis of the eigenvalue equation $\boldsymbol{\omega}_0 \cdot \nabla_{{\mathbf r}_0}Y({\mathbf r}_0,\boldsymbol{\omega}_0 )-\sigma Y({\mathbf r}_0,\boldsymbol{\omega}_0 )+\frac{\sigma \nu_1}{\beta} C^*\lbrace Y \rbrace =0$ leads to
\begin{equation}
\cosh \left( L\sigma \sqrt{1-\frac{\nu_1}{\beta}} \right) + \frac{(\beta-\frac{\nu_1}{2})\sinh \left( L\sigma\sqrt{1-\frac{\nu_1}{\beta}} \right) }{\sqrt{\beta(\beta-\nu_1)}}=0,
\label{eq_beta}
\end{equation}
which closely resembles Eq.~\ref{eq_lambda}. Imposing $\beta=1$ not surprisingly yields again Eq.~\ref{critical_length}. In a previous paper we derived Eq.~\ref{critical_length} based on the analysis of the moment $\langle n_V^1 \rangle$ as a function of the system parameters~\cite{zdm_epl}.

The survival probabilities in Eq.~\ref{extinction_time_eq} and~\ref{extinction_gen_eq} can be integrated numerically: in Figs.~\ref{fig2} and~\ref{fig3} we compare the resulting curves (as a function of time or generations, respectively) with Monte Carlo simulations for different configurations. In particular, once the probabilities $p_k$ have been chosen, by varying the rod size $L$ it is possible to impose $L<L_c$ or $L>L_c$. In the former case, the survival probabilities converge to zero independent of the starting direction, as expected, whereas in the latter the survival probabilities saturate to an asymptotic value $S_\infty$ that depends on the starting point as well as on the initial direction of the walker. Observe that $S^+_t$ up to a time of the order of $\tau^+ \simeq |L-x_0|/v$ does not feel the effects of the boundaries, yet, and the same holds for $S^-_t$ up to a time of the order of $\tau^- \simeq x_0/v$. Therefore, we expect $S_t^+ \simeq S_t^-$ up to $\min(\tau^+,\tau^-)$. For the configuration where $L>L_c$, the asymptotic survival probability as a function of starting point and initial direction is displayed in Fig.~\ref{fig4}.

We examine then the statistics of the travelled length $\ell_V$ and number of visits $n_V$ via the moment equations~\ref{moment_kac_ell_L} and~\ref{moment_eq_n}, respectively. In particular, we consider here the average and second moment, which in most cases are sufficient to characterize the typical behaviour of the stochastic variables and their dispersion. The equations are given in~\ref{rod_equations}, for a single neutron initially emitted at $x_0$. We consider the same geometrical configuration as above, and take $V(x,\omega)=V(x)$ for $x\in [0,L]$, i.e., we measure lengths and collisions independent of the local direction of the particles. As for boundary conditions, leakages impose $\langle \ell_V^m \rangle^+_t(L)=0$ and $\langle \ell_V^m \rangle^-_t(0)=0$; similarly, $\langle n_V^m \rangle^+_n(L)=0$ and $\langle n_V^m \rangle^-_n(0)=0$. For $L>L_c$ the moments would diverge as time and generations increase. Therefore, we choose a configuration such that $L<L_c$ and in Fig.~\ref{fig5} and~\ref{fig6} we plot the evolution of the moments of the travelled length and number of visits, respectively. Observe that the moments depend on the starting point as well as on the initial direction of the neutron, and level off to an asymptotic value, which is the signature of the process going to ultimate extinction. Furthermore, again for $\langle \ell_V^m \rangle_t$ we have $\langle \ell_V^m \rangle_t^+ \simeq \langle \ell_V^m \rangle_t^-$ up to $\min(\tau^+,\tau^-)$.

Observe finally that the rod model has been widely adopted to describe, among others, gas dynamics or biological species migration~\cite{kac_telegraph, bacteria, velocity_jump, weiss_review, lorentz}, so that the results described in this Section concerning an application in reactor physics could perhaps be of interest in other areas of science as well.

\section{Summary and conclusions}
\label{conclusions}

In this paper we have examined the statistics of travelled lengths and number of collisions for branching exponential flights. Moment formulas have been derived by resorting to the backward Feynman-Kac formalism, based on a minimal number of simplifying hypotheses. Moreover, we have shown that this same formalism can be extended with slight modifications to the analysis of other physical observables, such as the survival probability. The proposed formulas have been compared to Monte Carlo simulations for an example of one-dimensional transport inspired by reactor physics, and an excellent agreement was found. A generalization to more complex transport problems would also be possible, by relaxing for instance the requirement on the isotropy of the scattering kernel~\cite{zdm_pre_operator} and by introducing energy and space dependent cross sections. We conclude by observing that, although throughout this paper we have used our knowledge of $\psi$ and $\phi$ so to assess the statistical properties of the observables $\ell_V$ and $n_V$, the inverse is also possible: for instance, knowledge of the average total travelled length in $V$ allows inferring the average particle flux in that volume, and similarly knowledge of the total number of visits to $V$ allow inferring the average collision density in that volume.

\appendix

\section{The Feynman-Kac equations for $Q_t$ and $W_t$}
\label{fk_length}

Consider a single walker initially at ${\mathbf r}_0,\boldsymbol{\omega}_0$ at observation time $t=0$. We want to write an equation for the moment generating function $Q_t= \langle e^{-s \ell_V(t)}\rangle$. Assume an observation time $t+dt$: this can be split into a first interval, from $0$ to $dt$, and then a second interval from $dt$ to $t+dt$. The only requirement is that the process is Markovian: after $dt$ the particle continues its path without memory of the past. We start by observing that in a vanishing small time interval $dt$, from the definition of the underlying process, only two mutually exclusive events are possible: either the particle does not interact with the medium, in which case the walker keeps going in the same direction by a space interval $d{\mathbf r}_0 = v \boldsymbol{\omega}_0 dt$, or the particle interacts, in which case the walker disappears and gives rise to $k$ descendants at the same position, with random directions $\boldsymbol{\omega}'_k$ obeying the same probability density $C$. From the definition of the cross section $\sigma$, the former event happens at a rate $1-\sigma v dt$, whereas the latter at a rate $\sigma v dt$. If no particles are emitted ($p_0$), the trajectory is terminated, and no further contribution is added to $\ell_V$. When $k \ge 1$ (identical) particles are generated, the probability that the contribution to the total travelled length coming from each walker adds up precisely to $\ell_V$ is given by the convolution of the probability that the first particle spends a length $\ell_1$, the second $\ell_2$, and the $k$-th a length $\ell_V-\ell_1-\ell_2 - \cdots$. In the transformed space, the convolution products amount to a simple product of generating functions. This simple argument leads to the equation
\begin{eqnarray}
Q_{t+dt}(s|{\mathbf r}_0,\boldsymbol{\omega}_0)=(1-\sigma v dt) e^{-s v V({\mathbf r}_0,\boldsymbol{\omega}_0) dt} Q_t(s|{\mathbf r}_0+d{\mathbf r}_0,\boldsymbol{\omega}_0) +\nonumber \\
\sigma v dt \left[p_0 + p_1 \langle Q_t(s|{\mathbf r}_0,\boldsymbol{\omega}'_1)  \rangle + p_2 \langle  Q_t(s|{\mathbf r}_0,\boldsymbol{\omega}'_{21})  Q_t(s|{\mathbf r}_0,\boldsymbol{\omega}'_{22})\rangle + \cdots \right],
\end{eqnarray}
where brackets denote expectation with respect to the random directions $\boldsymbol{\omega}'_k$. Now, if we suppose that the descendant directions are independent, the expectation of a product of random variables becomes the product of expectations, so that we are led to
\begin{equation}
Q_{t+dt}=(1-\sigma v dt) e^{-s v V({\mathbf r}_0,\boldsymbol{\omega}_0) dt} Q_t(s|{\mathbf r}_0+d{\mathbf r}_0,\boldsymbol{\omega}_0) +\sigma v dt G\left[ \langle  Q_t(s|{\mathbf r}_0,\boldsymbol{\omega}_0') \rangle \right],
\end{equation}
where $G[z]=p_0 + p_1 z + p_2 z^2 + \cdots$ is the generating function associated to $p_k$. Observe that the average over the random directions can be expressed in terms of the associated probability density as
\begin{equation}
\langle  Q_t(s|{\mathbf r}_0,\boldsymbol{\omega}_0') \rangle = \int C^*(\boldsymbol{\omega}'_0 \to \boldsymbol{\omega}_0 |{\mathbf r}_0) Q_t(s|{\mathbf r}_0,\boldsymbol{\omega}'_0) d\boldsymbol{\omega}'_0,
\end{equation}
where $C^*$ is formally the adjoint density with respect to $C$. As a shorthand, we will denote $C^*\lbrace Q_t \rbrace = \int C^*(\boldsymbol{\omega}'_0 \to \boldsymbol{\omega}_0 |{\mathbf r}_0) Q_t(s|{\mathbf r}_0,\boldsymbol{\omega}'_0)d\boldsymbol{\omega}'_0$. Now, when $dt$ is small, at the leading order we have $Q_t(s|{\mathbf r}_0+d{\mathbf r}_0,\boldsymbol{\omega}_0) = Q_t(s|{\mathbf r}_0,\boldsymbol{\omega}_0)+ v \boldsymbol{\omega}_0 \cdot \nabla_{{\mathbf r}_0} Q_t dt + \cdots$, along the direction of $\boldsymbol{\omega}_0$. Furthermore, for vanishing $dt$ we have $\exp(-s v V({\mathbf r}_0,\boldsymbol{\omega}_0) dt)=1-svV({\mathbf r}_0,\boldsymbol{\omega}_0) dt + \cdots$. By recollecting all terms we then get
\begin{equation}
Q_{t+dt}= Q_{t} + v \boldsymbol{\omega}_0 \cdot \nabla_{{\mathbf r}_0} Q_t dt - v \sigma Q_t dt - s v V({\mathbf r}_0,\boldsymbol{\omega}_0) dt +\sigma v dt G\left[C^*\lbrace Q_t \rbrace \right].
\end{equation}
By dividing by $vdt$ and taking the limit $dt \to 0$, we finally obtain the backward Feynman-Kac equation for the moment generating function $Q_t$, namely
\begin{equation}
\frac{1}{v}\frac{\partial}{\partial t}Q_t=\boldsymbol{\omega}_0 \cdot \nabla_{{\mathbf r}_0}Q_t-\sigma Q_t - s V({\mathbf r}_0,\boldsymbol{\omega}_0 ) Q_t + \sigma G[C^*\lbrace Q_t \rbrace].
\end{equation}
Now, from the same argument as above, it follows that the probability generating function $W_t= \langle s^{m_V(t)}\rangle$ satisfies
\begin{eqnarray}
W_{t+dt}(s|{\mathbf r}_0,\boldsymbol{\omega}_0)=(1-\sigma v dt) W_t(s|{\mathbf r}_0+d{\mathbf r}_0,\boldsymbol{\omega}_0) +\nonumber \\
\sigma v dt \left[p_0 + p_1 \langle W_t(s|{\mathbf r}_0,\boldsymbol{\omega}'_1)  \rangle + p_2 \langle  W_t(s|{\mathbf r}_0,\boldsymbol{\omega}'_{21})  W_t(s|{\mathbf r}_0,\boldsymbol{\omega}'_{22})\rangle + \cdots \right],
\end{eqnarray}
hence
\begin{equation}
\frac{1}{v}\frac{\partial}{\partial t}W_t=\boldsymbol{\omega}_0 \cdot \nabla_{{\mathbf r}_0}W_t-\sigma W_t+ \sigma G[C^*\lbrace W_t \rbrace].
\end{equation}

\section{The discrete Feynman-Kac equation for $Q_n$}
\label{fk_visits}

Let a single walker be initially at ${\mathbf r}_0,\boldsymbol{\omega}_0$. When considering generations, it is more convenient to begin with a single particle entering its first collision with coordinates ${\mathbf r}_1,\boldsymbol{\omega}_0$, and we denote by $\tilde{Q}_{n}(u|{\mathbf r}_1,\boldsymbol{\omega}_0)=  \langle e^{-u n_V(n)}\rangle({\mathbf r}_1,\boldsymbol{\omega}_0)$ the corresponding moment generating function. The separation between ${\mathbf r}_1$ and ${\mathbf r}_0$, at a first glance somewhat artificial, is actually due to the special role of the source: a particle emitted from the source is just transported to the first collision point, and can not be absorbed nor multiplied at ${\mathbf r}_0$~\cite{zdm_pre_disc_kac, zdm_epl}. When entering the collision at ${\mathbf r}_1$, $k$ particles are created, with probability $p_k$, and random directions. Exponential flights are Markovian at collision points, which allows splitting each of the $k$ subsequent trajectories into a first jump, from ${\mathbf r}_1 $ to ${\mathbf r}'_{k}$ in direction $\boldsymbol{\omega}'_k$, and a branching path from ${\mathbf r}'_{k}$ to the positions held at the $(n+1)$-th generation. The displacement ${\mathbf r}'_{k}-{\mathbf r}_1$ obeys the jump length density $T$, and the direction $\boldsymbol{\omega}'_k$ the density $C$. If $k=0$, the trajectory ends at ${\mathbf r}_1 $ and there will be no further events contributing to $n_V$. Hence, we have
\begin{eqnarray}
\tilde{Q}_{n+1}(u|{\mathbf r}_1,\boldsymbol{\omega}_0)=p_0e^{-u V({\mathbf r}_1,\boldsymbol{\omega}_0)}+p_1e^{-u V({\mathbf r}_1,\boldsymbol{\omega}_0)}\langle \tilde{Q}_n(u|{\mathbf r}'_{1},\boldsymbol{\omega}'_1)\rangle+\nonumber \\
+p_2 e^{-u V({\mathbf r}_1,\boldsymbol{\omega}_0)}\langle \tilde{Q}_n(u|{\mathbf r}'_{21},\boldsymbol{\omega}'_{21})\tilde{Q}_n(u|{\mathbf r}'_{22},\boldsymbol{\omega}'_{22})\rangle + \cdots ,
\label{equation_F_one}
\end{eqnarray}
where expectation is taken with respect to the random displacements and directions, and the term $e^{-u V({\mathbf r}_1,\boldsymbol{\omega}_0)}$ can be singled out because it is not stochastic. The terms at the right hand side in Eq.~\ref{equation_F_one} can be understood as follows: the probability that $k$ identical and indistinguishable particles (born at ${\mathbf r}_1$) give rise to $n_V$ collisions in $V$ is given by the convolution product that the first makes $n_{1}$ collisions, the second $n_{2}$, ..., and the $k$-th $n_V-n_{1}-n_{2}-\cdots$. In the transformed space, this convolution becomes a simple product of generating functions. If we assume that the descendant particles are independent, the expectation of the products in Eq.~\ref{equation_F_one} becomes the product of the expectations. Now, observe that the average over displacements and directions can be expressed in terms of the associated densities, namely
\begin{equation}
\langle  \tilde{Q}_n(u|{\mathbf r}'_k,\boldsymbol{\omega}'_k) \rangle = \int C^*(\boldsymbol{\omega}'_0 \to \boldsymbol{\omega}_0 |{\mathbf r}_1) \int T^*({\mathbf r}'_1 \to {\mathbf r}_1 |\boldsymbol{\omega}'_0) \tilde{Q}_n(u|{\mathbf r}'_1,\boldsymbol{\omega}'_0) d{\mathbf r}'_1 d\boldsymbol{\omega}'_0,
\end{equation}
where $T^*$ is the adjoint density associated to $T$~\cite{zdm_pre_disc_kac}. Intuitively, $T^*$ displaces the walker backward in time. Observe in particular that the first collision coordinates ${\mathbf r}_1,\boldsymbol{\omega}_0$ obey the probability density $T^*({\mathbf r}_1 \to {\mathbf r}_0 |\boldsymbol{\omega}_0)$, namely,
\begin{equation}
Q_n(u|{\mathbf r}_0,\boldsymbol{\omega}_0)=\int \tilde{Q}_n(u|{\mathbf r}_1,\boldsymbol{\omega}_0) T^*({\mathbf r}_1 \to {\mathbf r}_0 |\boldsymbol{\omega}_0)d{\mathbf r}_1.
\end{equation}
Therefore, by using $C^*\lbrace Q_n \rbrace = \int C^*(\boldsymbol{\omega}'_0 \to \boldsymbol{\omega}_0 |{\mathbf r}_1) Q_n(u|{\mathbf r}_1,\boldsymbol{\omega}'_0) d\boldsymbol{\omega}'_0$ as above, we obtain the discrete Feynman-Kac equation in integral form, namely
\begin{equation}
\tilde{Q}_{n+1}(u|{\mathbf r}_1,\boldsymbol{\omega}_0) =e^{-u V({\mathbf r}_1,\boldsymbol{\omega}_0)} G\left[ C^*\lbrace Q_n \rbrace \right].
\label{eq_discrete_fk_integral}
\end{equation}
Finally, by integrating over $T^*$ both sides of Eq.~\ref{eq_discrete_fk_integral} we get
\begin{equation}
Q_{n+1}(u|{\mathbf r}_0,\boldsymbol{\omega}_0)
=\int d{\mathbf r}_1 T^*({\mathbf r}_1 \to {\mathbf r}_0 |\boldsymbol{\omega}_0) e^{-u V({\mathbf r}_1,\boldsymbol{\omega}_0)} G\left[ C^*\lbrace Q_n \rbrace \right].
\end{equation}
It can be shown~\cite{bell} that integral equations in the form
\begin{equation}
f({\mathbf r}_0,\boldsymbol{\omega}_0)=\int d{\mathbf r}_1 T^*({\mathbf r}_1 \to {\mathbf r}_0 |\boldsymbol{\omega}_0) g({\mathbf r}_1 , \boldsymbol{\omega}_0)
\end{equation}
can be equivalently recast into
\begin{equation}
\boldsymbol{\omega}_0 \cdot \nabla_{{\mathbf r}_0}f({\mathbf r}_0,\boldsymbol{\omega}_0) - \sigma f({\mathbf r}_0,\boldsymbol{\omega}_0) + \sigma g({\mathbf r}_0 , \boldsymbol{\omega}_0) = 0.
\end{equation}
Therefore, Eq.~\ref{eq_discrete_fk_integral} gives the discrete Feynman-Kac equation in integro-differential form
\begin{equation}
-\boldsymbol{\omega}_0 \cdot \nabla_{{\mathbf r}_0} Q_{n+1}(u|{\mathbf r}_0,\boldsymbol{\omega}_0) + \sigma Q_{n+1}(u|{\mathbf r}_0,\boldsymbol{\omega}_0) = \sigma e^{-u V({\mathbf r}_0,\boldsymbol{\omega}_0)} G\left[ C^*\lbrace Q_n \rbrace \right].
\end{equation}
Now, from the same argument as above, it follows that the probability generating function $W_n= \langle u^{m_V(n)}\rangle$ satisfies
\begin{equation}
\tilde{W}_{n+1}(u|{\mathbf r}_1,\boldsymbol{\omega}_0)=p_0+p_1\langle \tilde{W}_n(u|{\mathbf r}'_{1},\boldsymbol{\omega}'_1)\rangle
+p_2\langle \tilde{W}_n(u|{\mathbf r}'_{21},\boldsymbol{\omega}'_{21})\tilde{W}_n(u|{\mathbf r}'_{22},\boldsymbol{\omega}'_{22})\rangle + \cdots ,
\end{equation}
with $\tilde{W}_{n}({\mathbf r}_1,\boldsymbol{\omega}_0)=  \langle u^{m_V(n)}\rangle({\mathbf r}_1,\boldsymbol{\omega}_0)$. Finally,
\begin{equation}
-\boldsymbol{\omega}_0 \cdot \nabla_{{\mathbf r}_0} W_{n+1}(u|{\mathbf r}_0,\boldsymbol{\omega}_0) + \sigma W_{n+1}(u|{\mathbf r}_0,\boldsymbol{\omega}_0) = \sigma G\left[ C^*\lbrace W_n \rbrace \right].
\end{equation}

\section{The rod model equations}
\label{rod_equations}

From Eq.~\ref{moment_kac_ell_L}, the average travelled length obeys
\begin{equation}
\frac{1}{v}\frac{\partial}{\partial t}\langle \ell_V^1 \rangle_t ={\cal L}^*\langle \ell_V^1 \rangle_t + V(x_0 ) \langle \ell_V^{0} \rangle_t,
\end{equation}
which gives then
\begin{eqnarray}
\frac{1}{v}\frac{\partial}{\partial t}\langle \ell_V^1 \rangle^+_t = \frac{\partial}{\partial x_0}\langle \ell_V^1 \rangle^+_t -\sigma \langle \ell_V^1 \rangle^+_t + \frac{\sigma \nu_1 }{2}\left(\langle \ell_V^1 \rangle^+_t + \langle \ell_V^1 \rangle^-_t \right) + V(x_0 ) ,\nonumber \\
\frac{1}{v}\frac{\partial}{\partial t}\langle \ell_V^1 \rangle^-_t = -\frac{\partial}{\partial x_0}\langle \ell_V^1 \rangle^-_t -\sigma \langle \ell_V^1 \rangle^-_t + \frac{\sigma \nu_1 }{2}\left(\langle \ell_V^1 \rangle^+_t + \langle \ell_V^1 \rangle^-_t \right) + V(x_0 ).
\end{eqnarray}
For the second moment of the travelled length we have
\begin{equation}
\frac{1}{v}\frac{\partial}{\partial t}\langle \ell_V^2 \rangle_t ={\cal L}^*\langle \ell_V^2 \rangle_t + 2 V(x_0 )\langle \ell_V^{1} \rangle_t +  \sigma \nu_2 C^*\lbrace \langle \ell_V^1 \rangle_t \rbrace ^2,
\end{equation}
which gives
\begin{eqnarray}
\frac{1}{v}\frac{\partial}{\partial t}\langle \ell_V^2 \rangle^+_t = \frac{\partial}{\partial x_0}\langle \ell_V^2 \rangle^+_t -\sigma \langle \ell_V^2 \rangle^+_t + \frac{\sigma \nu_1 }{2}\left(\langle \ell_V^2 \rangle^+_t + \langle \ell_V^2 \rangle^-_t \right) \nonumber \\ + 2 V(x_0 )\langle \ell_V^{1} \rangle^+_t + \frac{\sigma \nu_2}{4} \left(\langle \ell_V^1 \rangle^+_t + \langle \ell_V^1 \rangle^-_t \right)^2,\nonumber \\
\frac{1}{v}\frac{\partial}{\partial t}\langle \ell_V^2 \rangle^-_t = -\frac{\partial}{\partial x_0}\langle \ell_V^2 \rangle^-_t -\sigma \langle \ell_V^2 \rangle^-_t + \frac{\sigma \nu_1 }{2}\left(\langle \ell_V^2 \rangle^+_t + \langle \ell_V^2 \rangle^-_t \right) \nonumber \\ + 2 V(x_0 )\langle \ell_V^{1} \rangle^-_t + \frac{\sigma \nu_2}{4} \left(\langle \ell_V^1 \rangle^+_t + \langle \ell_V^1 \rangle^-_t \right)^2.
\end{eqnarray}
From Eq.~\ref{moment_eq_n}, the average number of visits obeys
\begin{equation}
-\boldsymbol{\omega}_0 \cdot \nabla_{{\mathbf r}_0}  \langle n_V^{1} \rangle_{n+1}+ \sigma \langle n_V^{1} \rangle_{n+1} =
\sigma \nu_1 C^*\lbrace \langle n_V^{1} \rangle_n \rbrace +V(x_0 ),
\end{equation}
which gives
\begin{eqnarray}
-\frac{\partial}{\partial x_0} \langle n_V^{1} \rangle^+_{n+1} + \sigma \langle n_V^{1} \rangle^+_{n+1} =\frac{\sigma \nu_1}{2} \left(\langle n_V^{1} \rangle^+_n + \langle n_V^{1} \rangle^-_n \right) +V(x_0 ),\nonumber \\
\frac{\partial}{\partial x_0} \langle n_V^{1} \rangle^-_{n+1} + \sigma \langle n_V^{1} \rangle^-_{n+1} = \frac{\sigma \nu_1}{2} \left(\langle n_V^{1} \rangle^+_n + \langle n_V^{1} \rangle^-_n \right) +V(x_0 ).
\end{eqnarray}
For the second moment of the number of visits we have
\begin{eqnarray}
-\boldsymbol{\omega}_0 \cdot \nabla_{{\mathbf r}_0}  \langle n_V^{2} \rangle_{n+1}
+ \sigma \langle n_V^{2} \rangle_{n+1} =
\sigma \nu_1 C^*\lbrace \langle n_V^{2} \rangle_n \rbrace
+ \sigma \nu_2 C^*\lbrace \langle n_V^{1} \rangle_n \rbrace^2 \nonumber \\
+ 2 \sigma \nu_1 V(x_0 )C^*\lbrace \langle n_V^{1} \rangle_n \rbrace + \sigma V(x_0 ),
\end{eqnarray}
which leads to
\begin{eqnarray}
-\frac{\partial}{\partial x_0} \langle n_V^{2} \rangle^+_{n+1} + \sigma \langle n_V^{2} \rangle^+_{n+1} = \frac{\sigma \nu_1}{2} \left(\langle n_V^{2} \rangle^+_n + \langle n_V^{2} \rangle^-_n \right) \nonumber \\
+ \frac{\sigma \nu_2}{4} \left(\langle n_V^1 \rangle^+_n + \langle n_V^1 \rangle^-_n \right)^2
+ \frac{\sigma \nu_1}{2} \left(\langle n_V^{1} \rangle^+_n + \langle n_V^{1} \rangle^-_n \right) + \sigma V(x_0 )\nonumber \\
\frac{\partial}{\partial x_0} \langle n_V^{2} \rangle^+_{n+1} + \sigma \langle n_V^{2} \rangle^-_{n+1} = \frac{\sigma \nu_1}{2} \left(\langle n_V^{2} \rangle^+_n + \langle n_V^{2} \rangle^-_n \right) \nonumber \\
+ \frac{\sigma \nu_2}{4} \left(\langle n_V^1 \rangle^+_n + \langle n_V^1 \rangle^-_n \right)^2
+ \frac{\sigma \nu_1}{2} \left(\langle n_V^{1} \rangle^+_n + \langle n_V^{1} \rangle^-_n \right) + \sigma V(x_0 ).
\end{eqnarray}

\end{document}